\documentclass[twocolumn,10pt,pre, aps,superscriptaddress,showpacs]{revtex4-1}

\usepackage{subfigure}

\usepackage{dcolumn}
\usepackage{amsmath,amssymb}
\usepackage{bm}
\usepackage{color}
\usepackage{latexsym}
\usepackage{psfrag,graphicx}
\usepackage{color}
\usepackage[english]{babel}
\usepackage{latexsym}
\usepackage{epsf} 
\usepackage{amsfonts}
\usepackage{natbib}
\usepackage{epstopdf}
\usepackage{appendix}
\usepackage{changes}
\usepackage{siunitx}
\usepackage{float}

\definecolor{mygrey}{gray}{0.35}
\definecolor{myblue}{rgb}{0.2,0.2,0.8}
\definecolor{myzard}{cmyk}{0,0,0.05,0}
\definecolor{mywhite}{rgb}{1,1,1}
\definecolor{myred}{rgb}{1,0.,0.3}
\definecolor{MATblue}{rgb}{0 0.4470 0.7410}
\definecolor{MATorange}{rgb}{0.8477 0.3242 0.0977}
\definecolor{MATgreen}{rgb}{0 .75 0}

\usepackage[colorlinks=true,citecolor=myblue,linkcolor=myred]{hyperref}

\begin{document}

\title{Laser-plasma proton acceleration with a combined gas-foil target}

\author{Dan Levy}
\affiliation{Department of Physics of Complex Systems, Weizmann Institute of Science, Rehovot 76100, Israel}

\author{Constantin Bernert}
\author{Martin Rehwald}
\affiliation{Helmholtz-Zentrum Dresden-Rossendorf, Institute of Radiation Physics, Bautzner Landstr. 400, 01328 Dresden, Germany}
\affiliation{Technische Universit{\"a}t Dresden, 01069, Dresden, Germany}

\author{Igor A. Andriyash}
\thanks{Current address: Laboratoire d'Optique Appliqu\'{e}, ENSTA, Chemin de la Huni\`{e}re, 91761 Palaiseau, France}
\affiliation{Department of Physics of Complex Systems, Weizmann Institute of Science, Rehovot 76100, Israel}

\author{Stefan Assenbaum}
\affiliation{Helmholtz-Zentrum Dresden-Rossendorf, Institute of Radiation Physics, Bautzner Landstr. 400, 01328 Dresden, Germany}
\affiliation{Technische Universit{\"a}t Dresden, 01069, Dresden, Germany}

\author{Thomas Kluge}
\affiliation{Helmholtz-Zentrum Dresden-Rossendorf, Institute of Radiation Physics, Bautzner Landstr. 400, 01328 Dresden, Germany}

\author{Eyal Kroupp}
\affiliation{Department of Physics of Complex Systems, Weizmann Institute of Science, Rehovot 76100, Israel}

\author{Lieselotte Obst-Huebl}
\thanks{Current address: Lawrence Berkeley National Laboratory, Berkeley, California
94720, USA}
\affiliation{Helmholtz-Zentrum Dresden-Rossendorf, Institute of Radiation Physics, Bautzner Landstr. 400, 01328 Dresden, Germany}
\affiliation{Technische Universit{\"a}t Dresden, 01069, Dresden, Germany}

\author{Richard Pausch}
\affiliation{Helmholtz-Zentrum Dresden-Rossendorf, Institute of Radiation Physics, Bautzner Landstr. 400, 01328 Dresden, Germany}

\author{Alexander Schultze-Makuch}
\affiliation{Helmholtz-Zentrum Dresden-Rossendorf, Institute of Radiation Physics, Bautzner Landstr. 400, 01328 Dresden, Germany}
\affiliation{Technische Universit{\"a}t Dresden, 01069, Dresden, Germany}

\author{Karl Zeil}
\author{Ulrich Schramm}
\affiliation{Helmholtz-Zentrum Dresden-Rossendorf, Institute of Radiation Physics, Bautzner Landstr. 400, 01328 Dresden, Germany}
\affiliation{Technische Universit{\"a}t Dresden, 01069, Dresden, Germany}

\author{Victor Malka}
\affiliation{Department of Physics of Complex Systems, Weizmann Institute of Science, Rehovot 76100, Israel}

\begin{abstract}
    Laser-plasma proton acceleration was investigated in the Target Normal Sheath Acceleration (TNSA) regime using a novel gas-foil target. The target is designed for reaching higher laser intensity at the foil plane owing to relativistic self-focusing and self compression of the pulse in the gas layer. Numerical 3D particle-in-cell (PIC) simulations were used to study pulse propagation in the gas, showing a nearly seven-fold increase in peak intensity. In the experiment, maximum proton energies showed high dependence on the energy transmission of the laser through the gas and a lesser dependence on the size and shape of the pulse. At high gas densities, laser energy depletion and pulse distortion suppressed proton energies. At low densities, self-focusing was observed and comparable or higher proton energies were measured with the gas.
\end{abstract}

\maketitle

\section{Introduction} \label{sec:intro}
Laser-plasma ion acceleration is commonly accomplished using a thin $\sim 1\,\rm\mu m$ foil as a target. In this scheme, ions are accelerated owing to the well-explored Target Normal Sheath Acceleration (TNSA) mechanism \cite{wilks_energetic_2001}. For nearly two decades, TNSA with thin foils has demonstrated its robustness and ease of experimental realization. Within the TNSA regime, many different targets have been theorized and demonstrated, aiming for improving ion beam parameters over the simple thin foil target. These include coating the foil with foams \cite{passoni_energetic_2014,prencipe_development_2016, bin_ion_2015, bin_enhanced_2018, ma_laser_2019}, nanospheres \cite{margarone_laser-driven_2012}, micropillar arrays \cite{khaghani_enhancing_2017}, microchannels \cite{zou_laser-driven_2017} and even bacteria \cite{dalui_bacterial_2014}. In another approach, the foil is pre-irradiated by a weaker pulse, creating a plasma density gradient which can be controlled by the delay between the main and pre-pulse \cite{sentoku_high-energy_2002, lee_enhancement_2004, andreev_effect_2006, seo_effects_2007, yogo_laser_2007, glinec_evolution_2008, nuter_influence_2008, mckenna_effects_2008, zheng_preplasma_2013, esirkepov_prepulse_2014}. These methods exhibit improved performance mainly due to enhanced laser absorption in the first near-critical density layer which eventually translates into higher ion energies.

Here we propose a new approach for coupling the laser to the plasma using a target with a unique density profile. The target consists of helium gas several hundred microns long followed by a thin $5\,\rm \mu m$ stainless steel foil. The structure of a thick underdense layer followed by a thin overdense layer holds several distinct features. Firstly, the underdense layer is tunable in density, allowing for a continuous parameter scan. Secondly, the gas layer can be optically probed, enabling diagnosis of the pulse before it reaches the foil. Thirdly, gas is conveniently refreshed for every shot, thereby not demanding complex target fabrication such as coating the foils with foams and nanomaterials.

\begin{figure*}
    \centering
    \includegraphics[width=\textwidth]{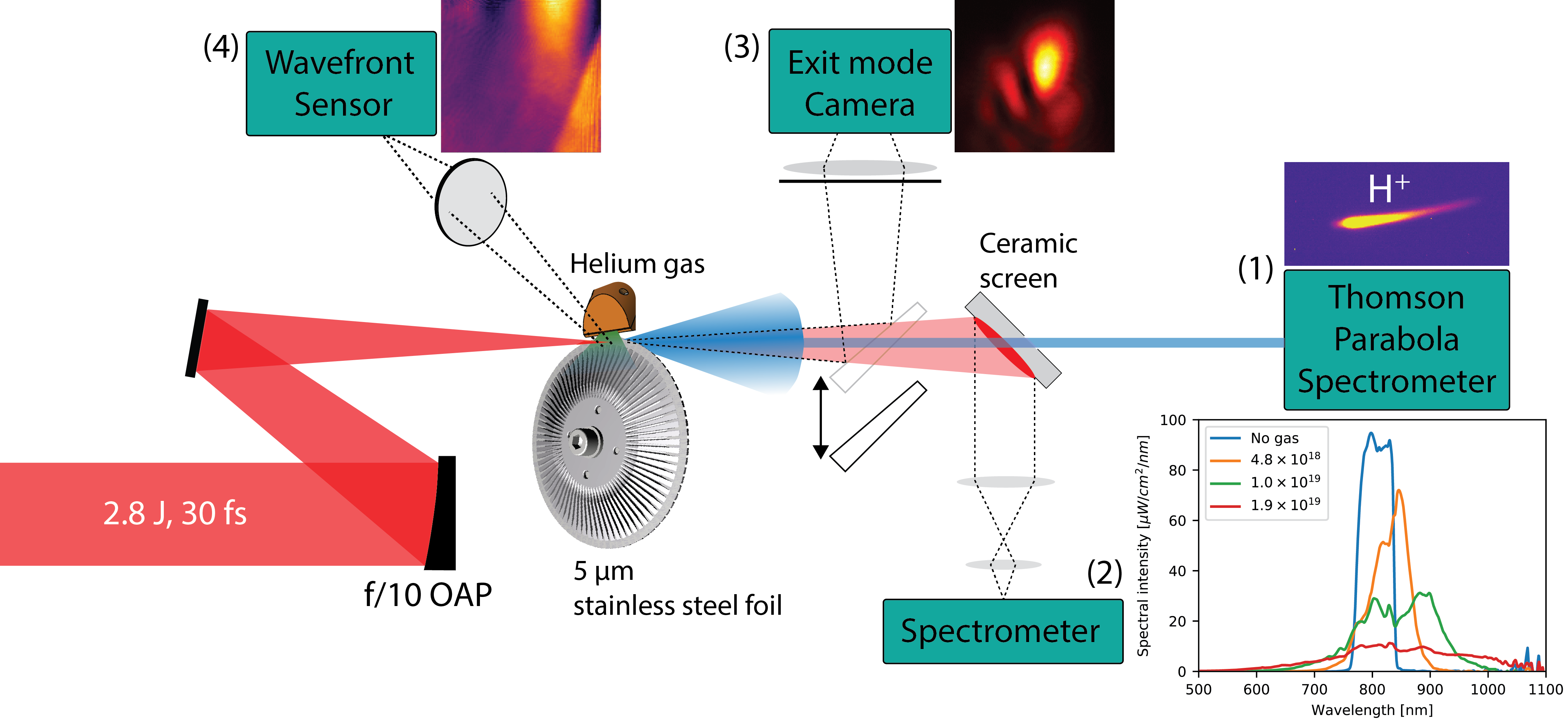}
    \caption{Experimental setup. The pulse impinges on the gas-foil target (helium gas, $5\,\rm \mu m$ stainless steel foil mounted on wheel). Diagnostics include: (1) a Thomson parabola charged particle spectrometer, (2) an optical spectrometer collecting light diffused by a ceramic screen placed after the target, (3) exit-mode imaging of the focal spot at the foil plane at full power, and (4), a wavefront sensor for on-line measurement of the gas density profile.}
    
    \label{fig:setup}
\end{figure*}

In TNSA, maximum proton energies were shown to scale as $(I_0\lambda^2)^\alpha$, where $I_0$ is peak laser intensity, $\lambda$ is the wavelength and $\alpha$ is a positive number \cite{zeil_scaling_2010, fuchs_laser-driven_2006, borghesi_fast_2006, robson_scaling_2007, flippo_scaling_2008}. The value of $\alpha$ depends on laser and target parameters and is generally about $0.5$. Moreover, $\alpha$ also depends on how $I_0$ is varied, whether by modifying pulse energy, duration or spot size \cite{esirkepov_laser_2006, schreiber_analytical_2006}.

As an ultraintense laser pulse propagates through an underdense plasma, it experiences relativistic self-focusing, self-compression and temporal steepening \cite{sprangle_relativistic_1987, faure_observation_2005, schreiber_complete_2010}. When the foil is placed at the right position, these effects potentially increase $I_0$ at the foil plane, thereby leading to enhanced proton energies. The pulse however also experiences energy depletion during propagation \cite{decker_evolution_1996}, diminishing the above positive effects. In a recent experiment, Streeter et al. demonstrated that under certain conditions temporal compression can dominate energy depletion, giving rise to an overall increase in peak power \cite{streeter_observation_2018}. Combining this power increase with the smaller self-focused spot thus has the potential of offering a significant boost in intensity after propagation in the plasma. In addition to the aforementioned effects, the pulse also undergoes spectral modulations during its propagation \cite{mori_physics_1997}. These shifts in frequency can happen in both the blue and red directions \cite{murphy_evidence_2006}. In case the red shift dominates the blue shift as in \cite{shiraishi_laser_2013}, the average increase in $\lambda$ could also be beneficial for ion acceleration.

Apart from the possibility of obtaining higher proton energies, the interaction of an ultrashort, ultraintense laser pulse with a gas-solid density profile provides a system of rich phenomena to explore. The study of this system was performed using multiple laser and particle diagnostics as well as PIC simulations. Diagnostics include imaging of the spatial shape of the pulse at the foil plane, laser spectrum and transmission measurements and a charged particle spectrometer. The ability to measure the pulse just before it reaches the foil is crucial in obtaining a better understanding of the underlying physics. By comparing proton measurements to laser measurements, the relation between the two can be inferred. In this work we present the study of such relations in the parameter space of the system.

\section{Experimental setup}
\label{sec:exp_setup}
\subsection{Laser and diagnostics}
\label{subsec:laser_and_diag}

The experimental setup is illustrated in figure \ref{fig:setup}. The DRACO $150\,\rm TW$ laser \cite{schramm_first_2017} with $2.8\,\rm J$ energy on target and a pulse duration of $30\,\rm fs$ is focused by an f/10 off-axis parabolic mirror (OAP) to $9.1\,\rm \mu m$ spot waist $w_0$. The $1\,\rm m$ focal length of the OAP enables positioning it with its back side facing the target in combination with a plain mirror to fold the beam. In this way, the expensive OAP does not suffer from any debris-induced damage. The auxiliary mirror is far from the target and is damaged at a slow rate. This is in contrast to common low f-number setups used for obtaining a small focal spot which require close proximity of the OAP to the target.

A Thomson parabola spectrometer was used for detection of accelerated charged particles. The particles enter through a $0.5\,\rm mm$ pinhole, get deflected by magnetic and electric fields and impact a Lanex scintillating screen placed perpendicular to the laser beam direction. Maximum kinetic energy $E_{max}$ of the particles can then be extracted from the resulting illuminated curves on the screen. The finite size of the pinhole introduces an uncertainty in this measurement. For protons, this uncertainty was calculated and found to be $\Delta E_{max} = 0.027\sqrt{E_{max}\,\rm[MeV]}$ for our geometry.

For transmitted laser spectrum detection, a $10\,\rm cm \times 10\,\rm cm$ ceramic screen is placed behind the target at $45^\circ$, collecting light transmitted through the gas. The light collection cone has a full angle of about $10^\circ$. The screen has a $3\,\rm mm$ hole in its center for transmission of particles from the target. The screen plane is imaged with two achromatic lenses onto a cosine corrector (optical diffuser). The corrector is connected to an intensity calibrated ensemble of a multi-mode fiber and an optical spectrometer for the detection of light between $300\,\rm nm$ and $1100\, \rm nm$. 

Relative transmission of laser light is calculated by integration of the spectrum curves. The spectral transfer function of all the optical elements between the screen and spectrometer was calculated and accounted for. Several measured curves for various densities are shown in Fig.~\ref{fig:setup}. The large fluctuations in spectral intensity seen above $1000\,\rm nm$ are a result of the amplification of the noise in a range where the spectrometer is less sensitive. This amplified noise introduces an uncertainty to the measurement.

For exit-mode imaging, a motorized wedge is inserted in the laser beam path, disabling particle and spectrum diagnostics. The foil plane is imaged with a 4" achromatic lens giving an effective spatial resolution of $7\,\rm \mu m$. The lens is preceded by two thin plastic foils coated by $40\,\rm nm$ of aluminum for further reducing fluence by nearly 3 orders of magnitude.

A wavefront sensor was used for in situ gas density measurements. Phase images integrated along the long side of the nozzle slit were obtained using a $532\,\rm nm$ laser diode for back-lighting (not shown in Fig.~\ref{fig:setup}). 

\subsection{The gas-foil target}
\label{subsec:gasfoil_target}

\begin{figure}[h!]
    \centering
    \includegraphics[width=\columnwidth]{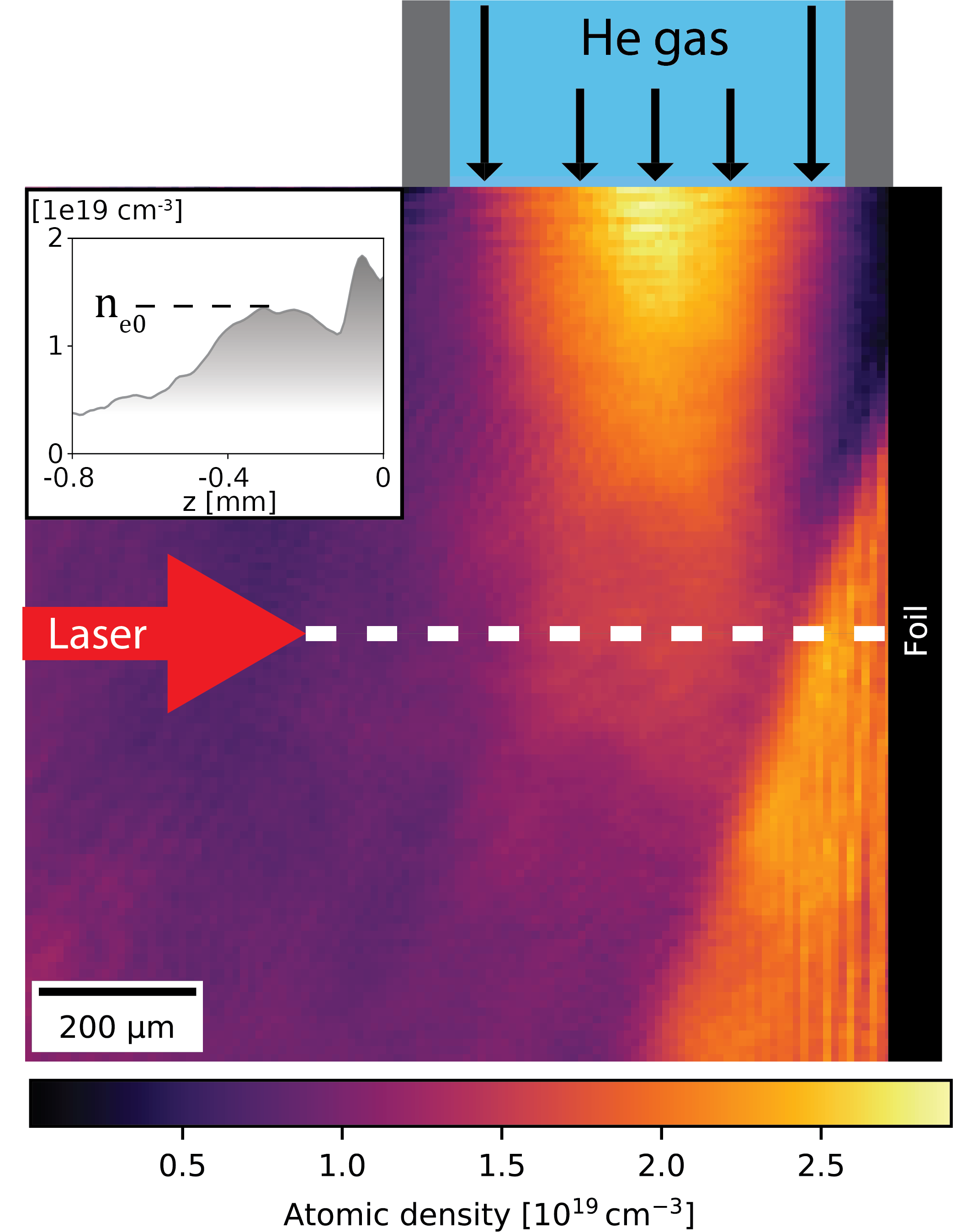}
    \caption{Calculated atomic density map from phase images taken with helium at 30 bar backing pressure. The nozzle is firing the gas downwards and the foil is placed close to its right edge. The laser is coming from the left, $0.5\,\rm mm$ below the end of the nozzle (dashed white line). The inset shows the gas density profile along this line.}
    \label{fig:SID4}
\end{figure}

The target consists of a helium gas exiting through a slit nozzle of dimensions $0.5\,\rm{mm} \times 5\,\rm{mm}$ followed by a $5\,\rm\mu m$ stainless steel foil, where the $0.5\,\rm mm$ side of the nozzle is perpendicular to the foil (see Fig. \ref{fig:setup}). The foil is precisely cut to fit on a rotating wheel. 

The nozzle slit width determines the length of the gas $L$ which the laser traverses before it impinges on the foil. Since absorption depends on 
the areal density of the underdense plasma column the laser interacts with $\int_{0}^{L}n_e(z)dz$, $L$ sets an upper limit on gas density: for a given length, there exists a density beyond which the laser energy will be completely absorbed in the gas. The narrower the nozzle slit, the higher this density is. Both relativistic self-focusing and self-compression increase with density, such that higher densities (but still undercritical) lead to smaller and shorter pulses \cite{mori_physics_1997}. Therefore, a thin and dense gas layer is generally preferable over a thick and dilute one. Experimentally we have found that gas flow was impeded for slit widths smaller than $0.5\,\rm mm$, which was hence the chosen size.

The slit geometry of a long rectangular shape was chosen for two reasons. First, it allows for estimation of the phase accumulation rate in the middle cross-section of the nozzle ($2.5\,\rm mm$). This is done by dividing the accumulated phase by the length of the nozzle, assuming planar symmetry. The gas density profile at this plane where the laser is fired is then deduced from this number. The second reason for choosing a long rectangular shape is that it ensures that the laser pulse interacts with a similar density profile for each shot despite laser pointing fluctuations.

A home-made gas valve was used in combination with an electronic pressure regulator. Pressure was increased up to $16\,\rm bar$ in the experiment. Steady-state density fluctuations were measured and found to be proportional to the inlet pressure, exhibiting an uncertainty of $7\%$. Measurements of density profiles between shots have shown good consistency when rotating the wheel and refreshing the foil. 

In Fig.~\ref{fig:SID4} we show the gas density map at the plane in which the laser propagates. The atomic density was calculated from measurements of the accumulated phase recorded by a wavefront sensor. The laser was fired $0.5\,\rm mm$ below the nozzle exit (white dashed line). The inset in the figure shows the density profile at this position. The foil induces a reflection of the gas, producing a narrow density peak adjacent to the foil. The density $n_{e0}$ is defined as the electron density of the wider peak further away from the foil, assuming full ionization of the helium gas. This value remains approximately the same regardless of whether the foil is in place or not. In the experiment, $n_{e0}$ was varied from $0$ (no gas) to $1.9\times10^{19}\,\rm cm^{-3}$ ($16\,\rm bar$).

\section{PIC simulations}
In order to model the laser-plasma interaction in the gas layer, 3D simulations were performed with the particle-in-cell PIConGPU~\cite{bussmann_radiative_2013} version 0.4.3 including updates~\cite{huebl_picongpu_2019}. The grid resolution was set to $177.2\,\rm nm\times 44.3\,\rm nm\times 177.2\,\rm nm$, with a PIC-cycle duration of $139\,\rm as$. The simulation box covered a volume of $512\times1536\times512$ cells. The interaction was modeled for more than $22000$ iterations using a moving window. The particle dynamics were computed using the Boris algorithm~\cite{boris_relativistic_1970}. Field updates and current deposition were handled by the Yee-solver~\cite{kane_yee_numerical_1966} and the Esirkepov current deposition scheme~\cite{esirkepov_exact_2001} using TSC particle shapes~\cite{hockney_computer_1988}. The entire simulation setup can be found online \cite{pausch_picongpu_2020}. 

The simulations were run for a $3\,\rm J$, $800\,\rm nm$ Gaussian pulse with $w_0=9.1\,\rm \mu m$ and a duration of $30\,\rm fs$ (FWHM). Peak $a_0$ in vacuum is 5.8, where $a_0=eA/m_ec^2\propto(I_0\lambda^2)^{0.5}$ is the normalized vector potential of the laser.

The plasma density was modeled using the measured gas profile as shown in Fig.~\ref{fig:SID4}, where $n_{e0}$ was varied from 0 to $2\times10^{19}\,\rm cm^{-3}$.

\begin{figure}
    \centering
    \includegraphics[width=\columnwidth]{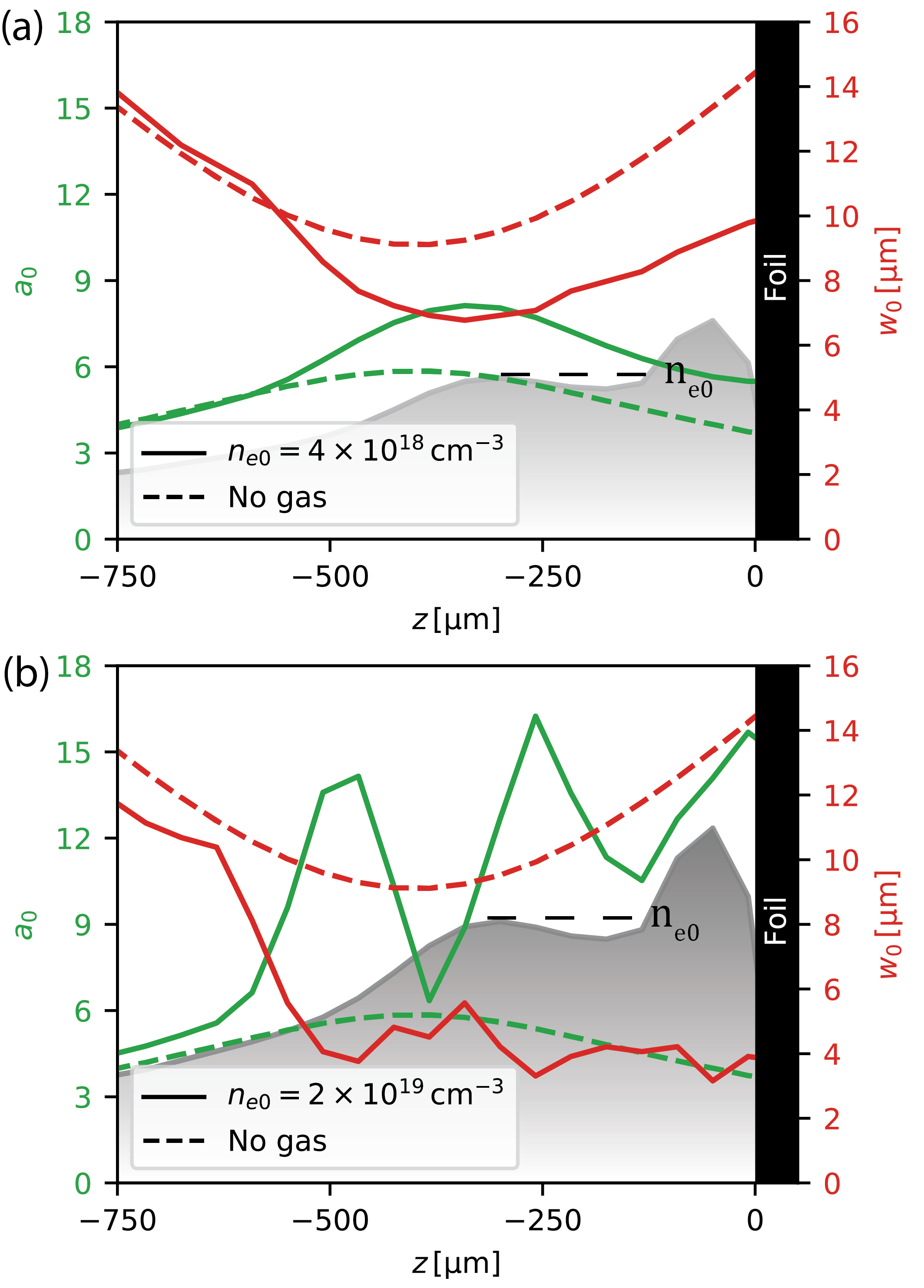}
    \caption{Comparison of $a_0(z)$ (green) and $w_0(z)$ (red) for simulations with $n_{e0}=4\times10^{18}\,\rm cm^{-3}$ (a) and $n_{e0}=2\times10^{19}\,\rm cm^{-3}$ (b). The dashed lines show the vacuum values (no gas). Vacuum focal plane is at $z=-400\,\rm \mu m$.}
    \label{fig:a0_w0_sim}
\end{figure}

Fig.~\ref{fig:a0_w0_sim} shows the evolution of $a_0$ and $w_0$ of the pulse as it propagates in the gas layer of low density $n_{e0}=4\times10^{18}\,\rm cm^{-3}$ (a) and high density $n_{e0}=2\times10^{19}\,\rm cm^{-3}$ (b). The shaded gray area represents the gas profile used for the simulation. Vacuum focal plane is at $z_{foc}=-400\,\rm \mu m$, about 1 Rayleigh length before the foil. We note that as the pulse propagates in the plasma its transverse shape is no longer Gaussian, mostly evident in the high density case. The values of $w_0$ are therefore calculated from FWHM slices ($w_0=0.85\,\rm FWHM$) and do not strictly represent a Gaussian beam.

In the low density case the effect of the gas is seen to be small. The pulse self-focuses down to about $7\,\rm \mu m$ and $a_0$ correspondingly increases to 8. The pulse then diffracts and reaches the foil plane with $a_0$ and $w_0$ similar to the no-gas values reached at $z_{foc}$. In the high density case, the interaction of the laser and the plasma is much stronger. The spot quickly self-focuses to about $w_0=4\,\rm \mu m$ and slightly oscillates around that value until reaching the foil. The $a_0$ curve shows three peaks of $\sim15$, reached at $z=-480\,\rm \mu m$, $z=-250\,\rm \mu m$ and $z=-10\,\rm \mu m$. The first peak is due to the drop in $w_0$. The second and the third peaks are due to the combined effects of the plasma on spot size, pulse duration and energy depletion. For this simulation, the pulse is compressed by a factor of 3 and about a quarter of its energy is lost by the time it reaches the foil.

The simulations show a slight red shift in average wavelength which increases with density, thereby having a small positive effect on $a_0$. In the low density case the average wavelength red shifts from $800\,\rm nm$ by $3\%$ by the time the pulse reaches the foil, whereas in the high density case it is shifted by $12\%$.

The value of $a_0$ at the foil plane $z=0$ is a non-trivial function of the two parameters $n_{e0}$ and $z_{foc}$. Since the strength of the interaction is governed by the plasma density, higher densities lead to earlier and stronger self-focusing, inducing oscillations in $a_0$ of larger amplitude. For a fixed $n_{e0}$, varying $z_{foc}$ roughly shifts $a_0(z)$ and can thus be used for reaching a peak at $z=0$.

Overall, simulations showed that an increase of a factor of about $2.5$ in $a_0$ (nearly 7 times in intensity) at the foil plane is possible using our laser parameters and measured gas density profile, as illustrated in Fig.~\ref{fig:a0_w0_sim}(b). This increase requires $n_{e0}$ to be in the range of $(1-2)\times10^{19}\,\rm cm^{-3}$. 

\section{Experimental results}

For a given laser pulse and gas density profile, pulse propagation is determined by the density of the gas $n_{e0}$ and the vacuum focal plane of the laser $z_{foc}$. We have experimentally explored this two-dimensional parameter space with the available particle and pulse diagnostics. For each setting of $n_{e0}$ and $z_{foc}$, a series of shots was performed: first, accelerated protons were measured. This shot leaves a $\sim1\,\rm mm$ hole in the foil. Then, more shots were taken in order to measure the effect of the gas on the pulse using the hole from the first shot. Only then the wheel was rotated and the foil was refreshed for the next measurement of protons. In this way, the laser conditions at the foil plane can be evaluated and compared to measured proton energies.

\begin{figure}
    \centering
    \includegraphics{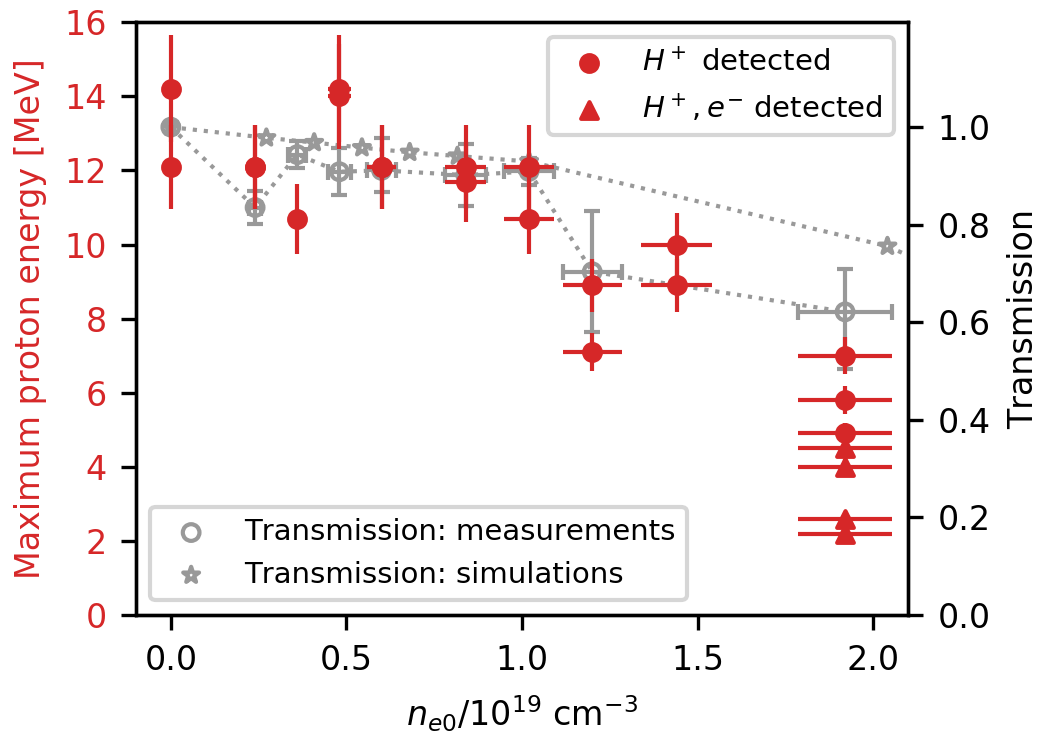}
    \caption{Maximum proton energies (red circles/triangles) as a function of $n_{e0}$ together with transmission values from measurements (gray circles) and PIC simulations (gray stars). Vacuum focal plane is fixed at $400\, \rm \mu m$ before the foil.}
    \label{fig:Ep_xmit}
\end{figure}

First, maximum proton energies were measured while scanning gas density with the vacuum focal plane fixed at $400\,\rm \mu m$ before the foil. The results are presented in Fig.~\ref{fig:Ep_xmit}. Protons were successfully accelerated when adding the gas, showing energies comparable to the no-gas case for up to $n_{e0}=1\times 10^{19}\,\rm cm^{-3}$. A decrease in proton energies is observed for higher densities. At $1.9\times10^{19}\,\rm cm^{-3}$ energetic electrons reaching up to $150\,\rm MeV$ were detected for some shots, together with  protons. These electrons originate in the gas and are accelerated by the wakefield created in the underdense plasma. The appearance of electrons only at some shots is likely a result of density fluctuations around the wavebreaking limit.

\begin{figure*}[ht!]
    \centering
    \includegraphics[width=\textwidth]{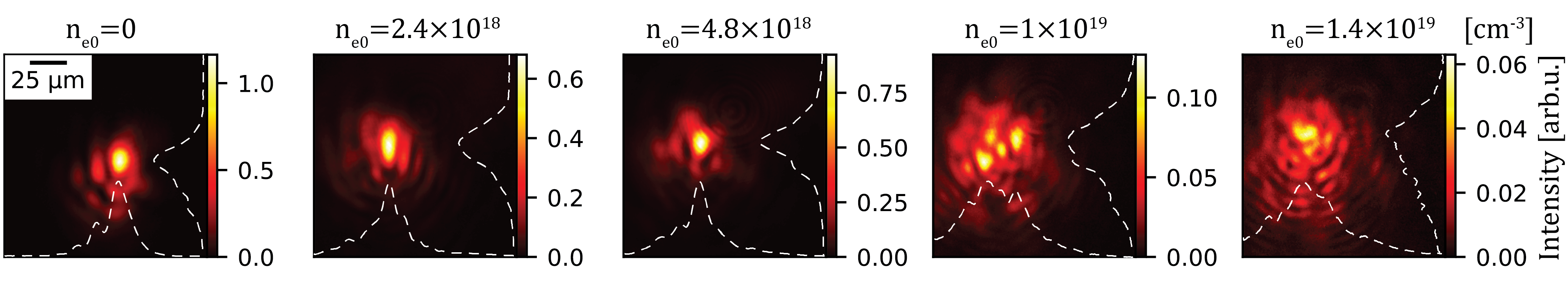}
    \caption{Focal spot images taken at the foil plane for different electron densities. The dashed white lines represent the projections of the images on the axes. Imaging was done through a hole in the foil.}
    \label{fig:exitmode}
\end{figure*}

In order to investigate the above behavior, both laser energy transmission and focal spot profile were measured at the foil plane. These measurements were done through a hole in the foil, as previously described. Transmission was measured by placing a ceramic screen behind the hole and imaging the resulting spot to a spectrometer. Alternatively, a wedge and a lens were placed in the beam path to image the focal spot at the hole plane ("exit mode imaging"). The measured spectra were also used for determining the shift in average wavelength. The averages were calculated by weighting according to the corresponding intensity curves and were seen to slightly red-shift with density, in agreement with simulations. At $1.9\times10^{19}\,\rm cm^{-3}$ the average wavelength shifted by about $7.5\%$ to $860\,\rm nm$. 

The transmission data are presented in Fig.~\ref{fig:Ep_xmit} alongside proton energies. The data points are the average of 2-4 measurements (shots) taken at the same conditions. The error bars show the highest and lowest measured values (measurement uncertainty included) of the corresponding data set. The transmission measurements are supplemented by simulations in order to validate their applicability.

We note that measured transmission values are correct for low densities $n_{e0}\lesssim10^{19}\,\rm cm^{-3}$, where at higher densities not all the transmitted light is collected due to light diffracting outside the ceramic screen boundaries. To account for this effect, images of the scattered light from the screen were recorded for each shot and a correction factor was applied accordingly. This factor was calculated by extrapolating the measured spot outside the screen assuming a circular shape around its center. The factor reached up to $1.18$. An uncertainty of $20\%$ in this factor is assumed, owing to the freedom in defining the spot center. The uncertainty in a single measurement is therefore calculated as the sum of the spectrometer noise and the error in the finite screen correction factor.

Simulations and measurements show good agreement below $1\times10^{19}\,\rm cm^{-3}$. The discrepancy above this density is possibly explained by a small fraction of the laser energy red-shifting beyond the spectrometer detection limit. Moreover, the pulse was observed to filament as shown further ahead. This beam breakup was not seen in simulations and could lead to increased depletion, thereby contributing to the discrepancy with the simulations.

Overall, maximum proton energies and transmission are seen to be highly correlated. This correlation suggests that proton energies are mostly sensitive to the pulse energy impinging on the foil. 

Fig.~\ref{fig:exitmode} shows measurements of spatial laser light intensity distributions ("focal spots") at the foil plane for 5 different densities. The three images for $n_{e0}=0$, $n_{e0}=2.4\times10^{18}\,\rm cm^{-3}$ and $n_{e0}=4.8\times10^{18}\,\rm cm^{-3}$ all show a distinct central spot, little changed by the addition of the gas. At the higher two densities, the spot is distorted and spreads over a large area, in contradiction with our simulations. At $n_{e0}=1\times10^{19}\,\rm cm^{-3}$, the deformation of the spot shape does not appear to affect proton energies which remain comparable to energies measured at lower densities with undistorted spots, as seen in Fig.~\ref{fig:Ep_xmit}. This further strengthens the hypothesis that pulse energy reaching the foil dominates the effect of the transverse spot profile in determining the maximum proton energy.

The observed beam breakup at high densities is possibly due to the collapse of the pulse within the plasma owing to side and forward Raman scattering instabilities \cite{antonsen_self-focusing_1992, ren_compressing_2001}. While we do not know where exactly in the plasma this breakup happens, it is possible that placing the foil at the peak of the first oscillation in $a_0$, i.e., at $z=-480\,\rm \mu m$ in Fig.~\ref{fig:a0_w0_sim} is essential for the scheme to work. This of course means that for our laser parameters a shorter gas layer should be used. 

\begin{figure}
    \centering
    \includegraphics[width=\columnwidth]{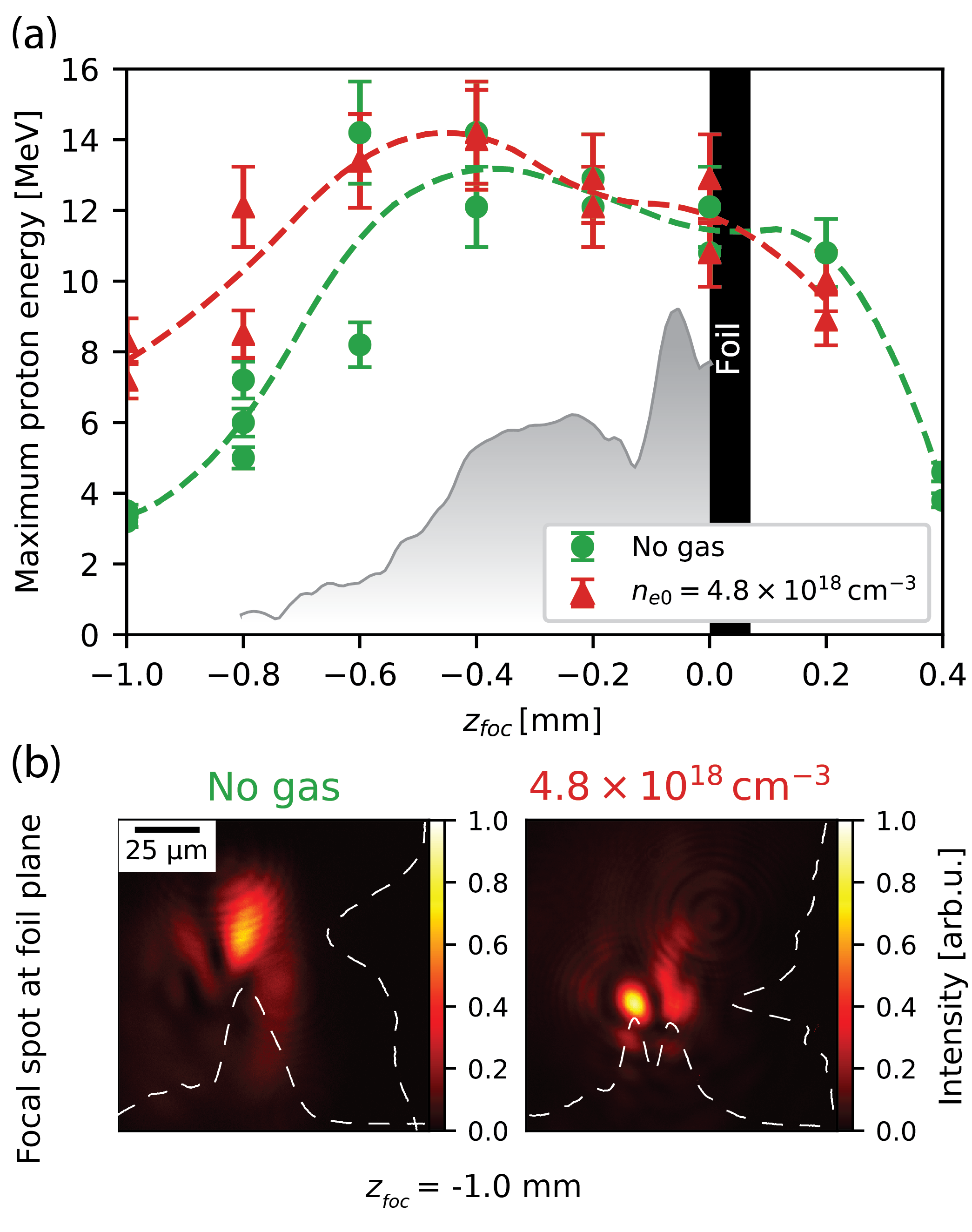}
    \caption{(a) Maximum proton energies as a function of vacuum focal plane position, with and without gas. The gas and foil are schematically shown. The dashed lines run through the averages of the shots and are for visualization purposes. (b) Focal spot images at the foil plane with and without gas, taken for $z_{foc}=-1\,\rm mm$. The white dashed lines represent the projections of the images on the axes.}
    \label{fig:zfoc_scan}
\end{figure}

The density scan has shown good pulse propagation at $n_{e0}=4.8\times10^{18}\,\rm cm^{-3}$ as well as high proton energies. Keeping the density fixed at this value, we have then varied the vacuum focal plane position $z_{foc}$ to see if reaching higher proton energies is possible in the vicinity of this well-behaved region in parameter space. While we did not detect an increase in overall maximum proton energy, a focal plane scan did however reveal a positive effect of the gas when the pulse is focused further away from the foil. In Fig.~\ref{fig:zfoc_scan}(a) maximum proton energies are shown as a function of $z_{foc}$ and are compared to the no-gas case. For $z_{foc}>-0.4\,\rm mm$, energies are comparable with/without gas. However when focusing further away from the foil, adding the gas is seen to be beneficial. At $z_{foc}=-1\,\rm mm$, the average maximum proton energy with the gas is more than twice higher than without the gas. 

Fig.~\ref{fig:zfoc_scan}(b) shows the focal spot images at the foil plane taken with and without the gas, for ${z_{foc}=-1.0\,\rm mm}$. In vacuum, a large bright spot appears at the top of a weak ring-like structure. This peculiar shape appears because the focal spot is optimized for $z_{foc}$. Since the beam is not Gaussian, the pulse changes its transverse shape as it diffracts. When the gas is added, a brighter and smaller circular spot appears at the center. We note that the reduction in spot size and increase in intensity were not so evident for all the shots where the gas showed enhanced proton energies. Moreover, for the same gas pressure and vacuum focal plane, different spots appeared when adding the gas, indicative of the non-linear nature of self-focusing. Therefore, the positive effect of the gas on proton energies is likely not due to self-focusing alone. 

\section{Summary}
\label{sec:summary}

Ions have been successfully accelerated for the first time using a novel gas-foil target. Proton energies correlated well with laser energy transmission through the underdense plasma layer and showed relatively little sensitivity to the spatial profile of the pulse. At low plasma densities of a few $10^{18}\,\rm cm^{-3}$, maximum proton energies were comparable to the no-gas case and the laser pulse propagated well in the underdense plasma. At high densities $(1-2)\times10^{19}\,\rm cm^{-3}$, where an increase in intensity at the foil plane was expected according to simulations, proton energies declined and the pulse appeared distorted when reaching the foil plane. The distortion of the pulse did not appear in our simulations, motivating further study of pulse propagation in the little-explored few-$10^{19}\,\rm cm^{-3}$ density regime. 

A focal plane scan showed that when the laser is focused far in front of the foil, adding the gas can enhance proton energies by more than twice. A reduction in spot size and increase in intensity was observed at the foil plane and serves as a partial explanation to the positive effect of the gas. 

Obtaining higher overall energies compared to bare foils demands a high energy transmission through the plasma due to the strong scaling of proton energies with laser energy. Additionally, it appears that the foil needs to be placed at the first intensity peak where the pulse is maximally focused, before it breaks up. Both of these conditions can be met by using a larger initial spot and/or a shorter gas layer. 

\section*{Acknowledgements}
We thank the DRACO laser team for their excellent support. The project was pursued in the framework of the Weizmann Helmholtz Laboratory for Laser Matter Interaction (WHELMI) and partially supported by Horizon 2020 Laserlab Europe V (PRISES) under contract no. 871124, as well as by The Israel Science Foundation (under contracts 666/17 and 711/17), Minerva (under contract 712590) and the Alexander von Humboldt Foundation.

\bibliography{bibliography}
\bibliographystyle{apsrev4-1}

\end{document}